\documentclass[12pt]{iopart}

\usepackage{iopams}
\usepackage[english]{babel}
\usepackage{graphicx}
\usepackage{cite}
\usepackage{bbm}
\usepackage{mathenv}
\usepackage{color}
\usepackage{float}

\begin{document}

\title{Structure invariant wave packets}

\author{V. Arrizon, F. Soto-Eguibar and H.M. Moya-Cessa}

\address{Instituto Nacional de Astrof\'{\i}sica, \'Optica y Electr\'onica,\\ Calle Luis Enrique Erro 1, Santa Mar\'{\i}a Tonantzintla, Puebla, 72840 Mexico}

\begin{abstract}
We show that by adding a quadratic phase to an initial arbitrary wavefunction, its free evolution maintains an invariant structure while it spreads by the action of an squeeze operator. Although such invariance is an approximation, we show that it matches perfectly the exact evolution.
\end{abstract}

%\pacs{42.50.Ct, 42.65.Ky, 03.65.2w}

%\ocis{ (050.5298) Photonic crystals;  (230.7370) Waveguides;  (350.5500) evolution.}

%\submitto{\PS}
\maketitle

\section{Introduction}
The Schr\"odinger equation for a free particle has attracted the search for wave functions that evolve without distortion.  Berry and Balasz have shown that an Airy wave function keeps its form under evolution, just showing some acceleration \cite{Berry}. However, Airy wave functions are not square integrable functions and therefore are not proper wave functions. If one wants to use  them, they need to be apodized, either by cutting them or by super-imposing a Gaussian function; i.e., instead considering a Gauss-Airy beam. In such case, it is too much to say that they loose their shape as they evolve, and therefore, their beauty. Effects such as focusing of waves may occur when particles go through a single slit \cite{Schleich2}, as it has been shown by studying the time dependent wave function in position space and its Wigner function \cite{Schleich1}.\\
In this contribution, we want to show that by adding a positive quadratic phase to an initial arbitrary wavefunction, its free evolution maintains an invariant structure, while it spreads by the action of an squeeze operator. That means, that the effect of passing a beam of particles (for instance electrons \cite{elec}, neutrons \cite{neut} or atoms \cite{atom}) through a negative lens, provides the wave function with the property of evolution invariance, while it diffracts by the application of a squeeze operator to the initial state \cite{Yuen,Caves,Satya,Vidiella,Knight,Schleich}.\\
In the following, we will revisit Airy beams and Airy-Gauss beams in order to show that the later ones deform as they evolve. In Section III, we show that the acquisition of a quadratic phase helps any field to become invariant under free evolution; in Section IV, we give some examples, namely initial  Sinc  and Bessel  functions, while Section V is left for conclusions.

\section{Revisiting Airy beams}
Berry and Balasz \cite{Berry} have shown that an initial wave function of the form (for simplicity we set $\hbar=1$)
\begin{equation}\label{Airy0}
\psi(x,0)=\mathrm{Ai}(\epsilon x),
\end{equation}
where $\epsilon$ is an arbitrary real constant, evolves according to the Schr\"odinger equation for a free particle of mass $m=1$
\begin{equation}\label{schr0}
i\frac{\partial \psi(x,t)}{\partial t}=\frac{\hat{p}^2}{2}\psi(x,t),
\end{equation}
as
\begin{equation}
\psi(x,t)=\mathrm{Ai}\left[\epsilon \left(x-\frac{\epsilon^3t^2}{4}\right)\right]
\exp\left[ i \frac{\epsilon^3t}{2}\left( x-\frac{\epsilon^3t^2}{6}\right)\right],
\end{equation}
as can be verified by substitution into (\ref{schr0}). It is clear from this solution that the Airy wave packet is conserved, meaning that it evolves without spreading. Besides, the evolution shows an acceleration which may be obtained also in some other initial distributions of wave packets, like half Bessel functions \cite{Aleahmad}. Propagation of Airy wavelet-related patterns has also been considered in \cite{Torre0} and it has been shown they provide \textquotedblleft source functions\textquotedblright \; for freely propagating  paraxial fields. The acceleration may be corrected by propagating the Airy function in a linear potential \cite{Chavez}. Unfortunately, the Airy wave packet is not a proper wave function as it is not square integrable. A possibility for making it normalizable would be to cut it (have a window) or to {\it apodize} it by multiplying it by a Gauss function, and effectively cutting it. If instead of  the initial state (\ref{Airy0}), we consider as initial condition the normalizable wave function
\begin{equation}\label{Airy1}
\psi(x,0)=\textrm{Ai}(\epsilon x)\exp\left( -\beta x^2\right),
\end{equation}
with $\beta$ another arbitrary real constant, the solution then reads
\begin{equation}\label{Airy1Sol}
\psi(x,t)=\frac{1}{\sqrt{1-2i\beta t}}\textrm{Ai}\left[ \zeta(x,t)\right] \exp\left( \frac{\beta x^2}{2i\beta t -1}\right) \exp\left[i\gamma(x,t)\right] 
\end{equation}
with
\begin{equation}
\zeta(x,t)=\frac{\epsilon^4t^2+\epsilon x(2i\beta t -1)}{(2\beta t+i)^2},   \qquad 
\gamma(x,t)=\frac{3\epsilon^3xt(2\beta t+i)-2i\epsilon^6t^3}{3(2\beta t+i)^3};
\end{equation}
again, this can be proved by direct substitution into Eq.(\ref{schr0}). In Figure 1, we plot the probability density $|\psi(x,t)|^2$ for Eq.(\ref{Airy1Sol}) for different times. We can see that for $\beta=0.01$, the Airy-Gauss beam still accelerates, but it looses its shape.
\begin{figure}[H]
\centering{}
\includegraphics[width=12cm]{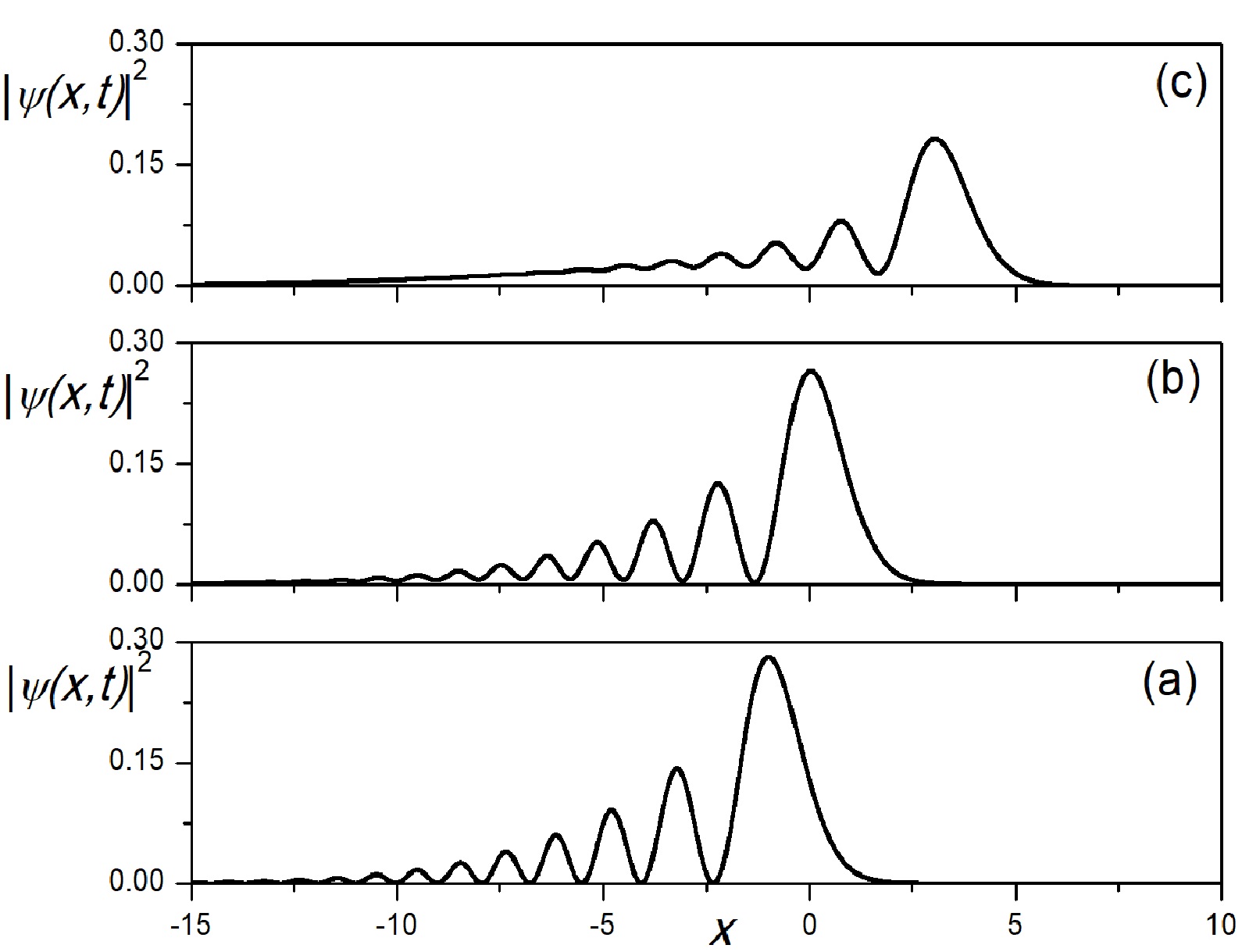} 
\caption{Plot of the probability density $|\psi(x,t)|^2$ of the wavefuntion in equation (\ref{Airy1Sol}) for the parameters $\epsilon=1$ and $\beta=0.01$ at (a) $t=0$, (b) $t=1$ and (c) $t=2$.}
\label{fig1}
\end{figure}

\section{Evolution invariant beams}
Now consider an initial condition of the form
\begin{equation}
\psi(x,0)=\exp\left( i\alpha x^2\right) \phi(x,0),
\end{equation}
where $\alpha$ is a real parameter which must be set in each specific case \cite{victor18}. The solution of the Schr\"odinger equation then reads
\begin{equation}\label{sol}
\psi(x,t)=\exp\left( -i\frac{t}{2}\hat{p}^2\right) \exp\left( i\alpha x^2\right)  \phi(x,0).
\end{equation}
Writing the identity operator as $\hat{I}=\left( i\frac{t}{2}\hat{p}^2\right) \exp\left( -i\frac{t}{2}\hat{p}^2\right)$, the previous equation can be cast as
\begin{equation}\label{ec090}
\psi(x,t)=\exp\left( -i\frac{t}{2}\hat{p}^2\right) 
\exp\left( i\alpha x^2\right)  
\exp \left( i\frac{t}{2}\hat{p}^2\right) \exp\left( -i\frac{t}{2}\hat{p}^2\right) \phi(x,0).
\end{equation}
As is well known, $\exp\left( -i\frac{t}{2}\hat{p}^2\right) x \exp\left( i\frac{t}{2}\hat{p}^2\right) =x-t\hat{p}$, and this implies that
\begin{eqnarray}
\exp\left( -i\frac{t}{2}\hat{p}^2\right) 
\exp\left( i\alpha x^2\right)  
\exp \left( i\frac{t}{2}\hat{p}^2\right)&=&\exp \left[ i \alpha \left( x-t\hat{p}\right) ^2 \right]
\\ \nonumber
&=&\exp\left\lbrace i\alpha [x^2-t(x\hat{p}+\hat{p}x)+t^2\hat{p}^2]\right\rbrace,
\end{eqnarray}
which substituted in equation (\ref{ec090}) gives us 
\begin{equation} 
\psi(x,t)=\exp\left\lbrace i\alpha [x^2-t(x\hat{p}+\hat{p}x)+t^2\hat{p}^2]\right\rbrace \exp\left( -i\frac{t}{2}\hat{p}^2\right) \phi(x,0).
\end{equation}
It is not difficult to show that the first exponential above may be factorized as \cite{metop}
\begin{equation} 
\exp \left[ if_1(t)x^2\right]  \exp\left[ i f_2(t)(x\hat{p}+\hat{p}x)\right] \exp\left[ i f_3(t)\hat{p}^2\right] ,
\end{equation}
with
\begin{equation}
f_1(t)= \frac{\alpha}{1+2\alpha t}, \qquad f_2(t)=-\frac{1}{2} \ln(1+2\alpha t), \qquad f_3(t)= \frac{\alpha t^2}{1+2\alpha t}.
\end{equation}
This allows us to give a final form for equation (\ref{sol}) as
\begin{equation}
\psi(x,t)=\exp\left[ if_1(t)x^2\right]  \exp\left[ if_2(t)(x\hat{p}+\hat{p}x)\right] \exp\left[ i f_4(t)\hat{p}^2\right] \phi(x,0) 
\end{equation}
with $f_4(t)=f_3(t)-t/2$.\\
We now examine the behaviour of $f_4(t)$ as a function of the parameter $\alpha$. The Taylor series of $f_4(t)$ for $\alpha \approx 0$ is
\begin{equation}\label{0150}
f_4(t) = -\frac{t}{2}+t^2 \alpha-2t^3 \alpha^2+\textrm{O}(\alpha)^3
\end{equation}
and for $\alpha \approx \infty$ is
\begin{equation}\label{0160}
f_4(t) = -\frac{1}{4\alpha}+\frac{1}{8t\alpha^2}+\textrm{O}\left( \frac{1}{\alpha}\right) ^3.
\end{equation}
In Figure 2, we plot  $f_4(t)$  as a function of time for different values of the $\alpha$ parameter. It may be seen that for small values of $\alpha$ it remains close to zero, and for large values of $\alpha$ it becomes very small, as expected from the approximation in Equation (16).
\begin{figure}[H]
\centering{}
\includegraphics[width=10cm]{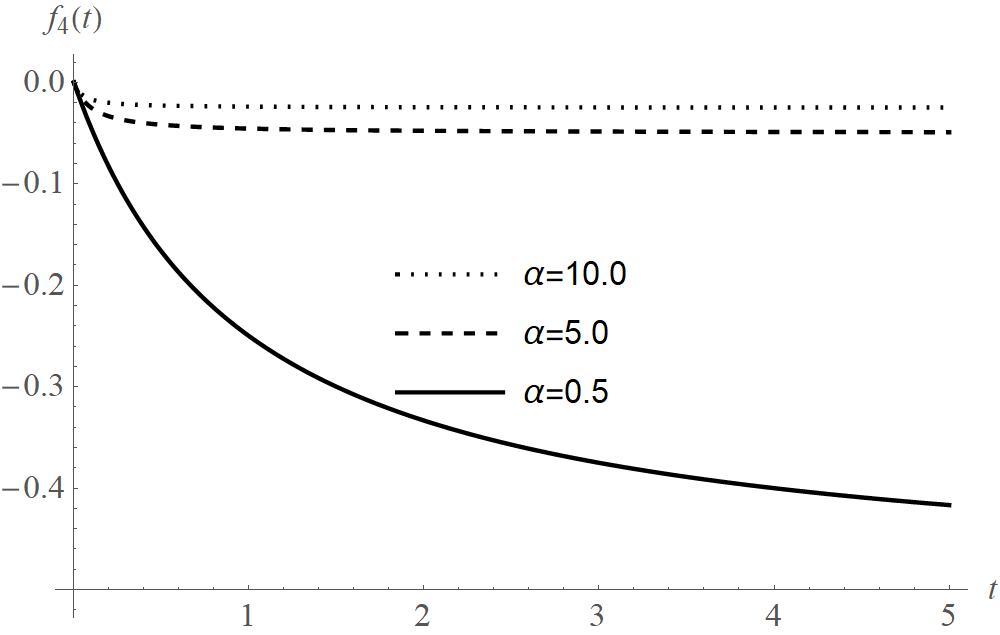} 
\caption{Plot of the function  $f_4(t)$ for  $\alpha=10$  (dotted line), $\alpha=5.0$ (dashed line) and $\alpha=0.5$ (continuous line).}
\label{fig2}
\end{figure}
Thus, for small values of $\alpha$, we take the first two terms in the Taylor development of the operator 
$\exp\left[ i f_4(t)\hat{p}^2\right]$ and we get
\begin{equation} \label{appsol}
\psi_1(x,t)\approx \exp\left[ if_1(t)x^2\right]  \exp\left[ if_2(t)(x\hat{p}+\hat{p}x)\right]  \left[1+if_4(t)\hat{p}^2\right]\phi(x,0).
\end{equation}
For $\alpha$ large enough, we completely disregard the term $\exp\left[ i f_4(t)p^2\right] $ and, to a very good approximation (as will be see below), we write simply the zeroth order solution
\begin{equation}\label{appsol2}
\psi_0(x,t)\approx \exp\left[ if_1(t)x^2\right]  \exp\left[ i f_2(t)(x\hat{p}+\hat{p}x)\right] \phi(x,0).
\end{equation}
The operator $\exp\left[i f_2(t)(x\hat{p}+\hat{p}x)\right]$ is the squeeze operator, and by its application to the initial function, the equation above may be cast into
\begin{equation}\label{appsol3}
\psi_0(x,t)=\frac{1}{\sqrt{1+2\alpha t}} \exp\left[ if_1(t)x^2\right]  \phi\left(\frac{x}{1+2\alpha t},0\right). 
\end{equation}
It is clear that the above wave function gives a probability density that remains invariant during evolution
\begin{equation}\label{invariant}
|\psi_0(x,t)|^2=\frac{1}{{1+2\alpha t}}\left\vert \phi\left(\frac{x}{1+2\alpha t},0\right)\right\vert^2.
\end{equation}
The choice of the $\alpha$ parameter depends of the problem that is being studied and on the propagation distance that must be considered, as will be shown in the examples below. From Eqs. (\ref{0150}) and (\ref{0160}), it is also clear that different values of $\alpha$ must be considered if the zeroth order or the first order solutions are going to be used. In  \cite{victor18} we present a discussion on the election of this parameter in the realm of classical optics.

\section{Some examples}
In this section, we study some examples where we apply our approximation and compare it with the exact solution. 
\subsection{Sinc function}
\begin{figure}[H]
\centering{}
\includegraphics[width=11cm]{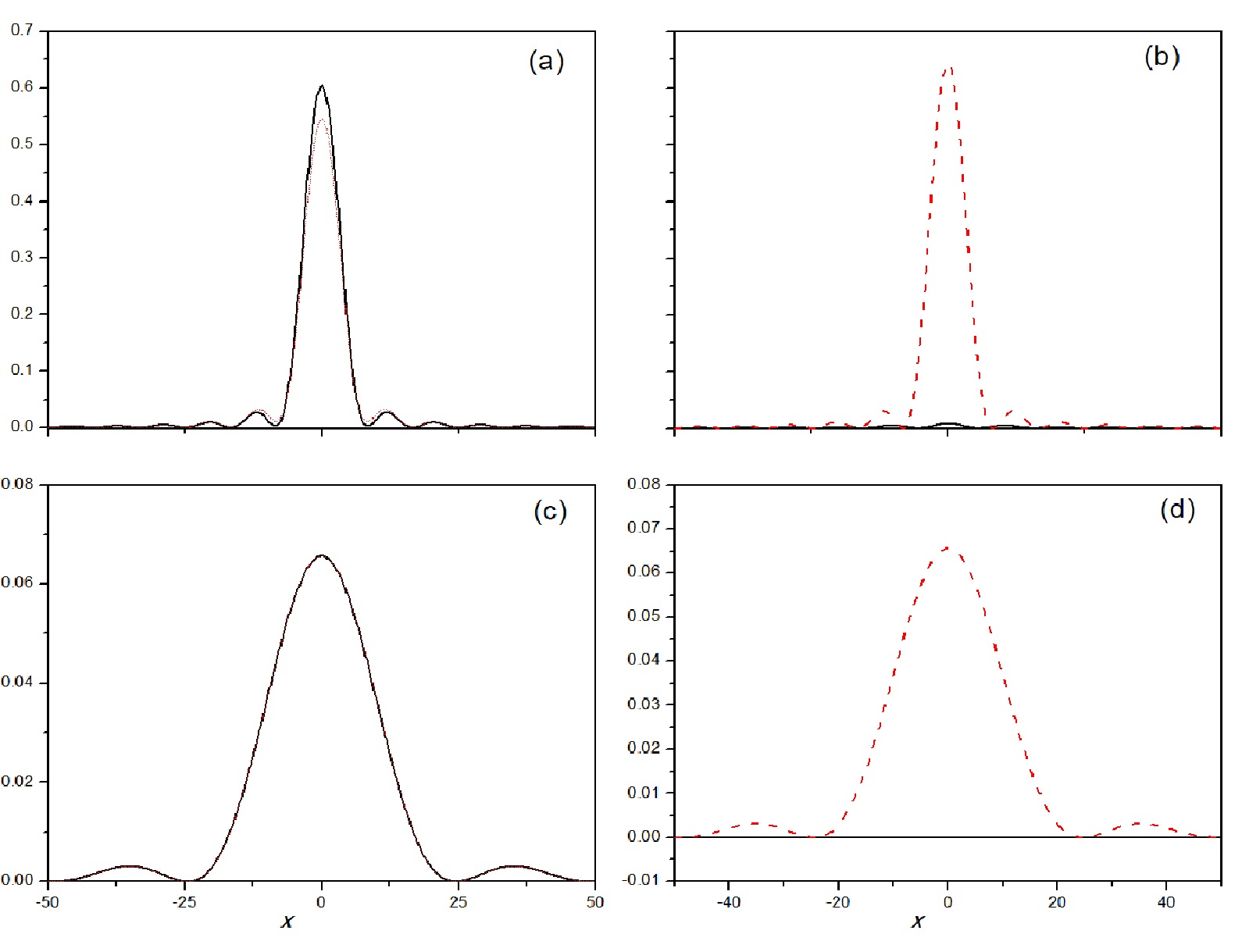} \caption{\label{fig3} Plot of the probability densities for an initial Sinc function as given in equation (\ref{sinc}) with $b=1$ and the parameters (a) $\alpha=0.3$ and (c) $\alpha=3$ for $t=5$ for the exact (solid line) and approximate solutions (dashed line). In  (b) and (d) are the comparison between the zero (dashed line) and first order (solid line) contributions.}
\end{figure}

We start with an initial (unnormalized, but normalizable) wave packet of the form
\begin{equation} 
\psi(x,0)= \exp\left( i\alpha x^2\right)  \mathrm{Sinc}(bx),
\end{equation}
where $b$ is an arbitrary real constant and where we define the Sinc function as
\begin{equation} 
\mathrm{Sinc}(bx)=\frac{1}{b}\int_{-b}^b \exp\left( i u x\right)  du.
\label{sinc}
\end{equation}
We write the approximations to zeroth and first order as
\begin{equation} 
\psi_0(x,t)=\frac{\exp\left[ if_1(t)x^2\right]} {b\sqrt{1+2\alpha t}} \int_{-b}^b \exp\left( iu\frac{x}{1+2\alpha t}\right) du,
\end{equation}
and
\begin{equation} 
\psi_1(x,t)=\psi_0(x,t)+i\frac{f_4(t) \exp\left[ if_1(t)x^2\right]} {b\sqrt{1+2\alpha t}} \int_{-b}^b u^2 \exp\left( iu\frac{x}{1+2\alpha t}\right) du,
\end{equation}
respectively. For the sake of comparison, we can  also write the exact solution as 
\begin{equation} 
\psi(x,t)=\frac{\exp\left[ if_1(t)x^2\right] }{b\sqrt{1+2\alpha t}} \int_{-b}^b \exp\left[ if_4(t)u^2\right]  \exp\left( iu\frac{x}{1+2\alpha t}\right) du.
\end{equation}
We plot in Figure \ref{fig3} (a) and (c) the probability densities for the zeroth order and exact solutions, showing that they match very well for a value of $\alpha=0.3$ and have an excellent agreement for a greater value ($\alpha=3$). In Figure \ref{fig3} (b) and (d), the quantities $|\psi_0(x,t)|^2$ (dashed line) and $|\psi_0(x,t)-\psi_0(x,t)|^2$ (solid line) are plotted in order to show that their contributions to the first order approximation are negligible, already for such small values of $\alpha$. 

\subsection{Bessel function}
\begin{figure}[H] 
\centering{}\includegraphics[width=12cm]{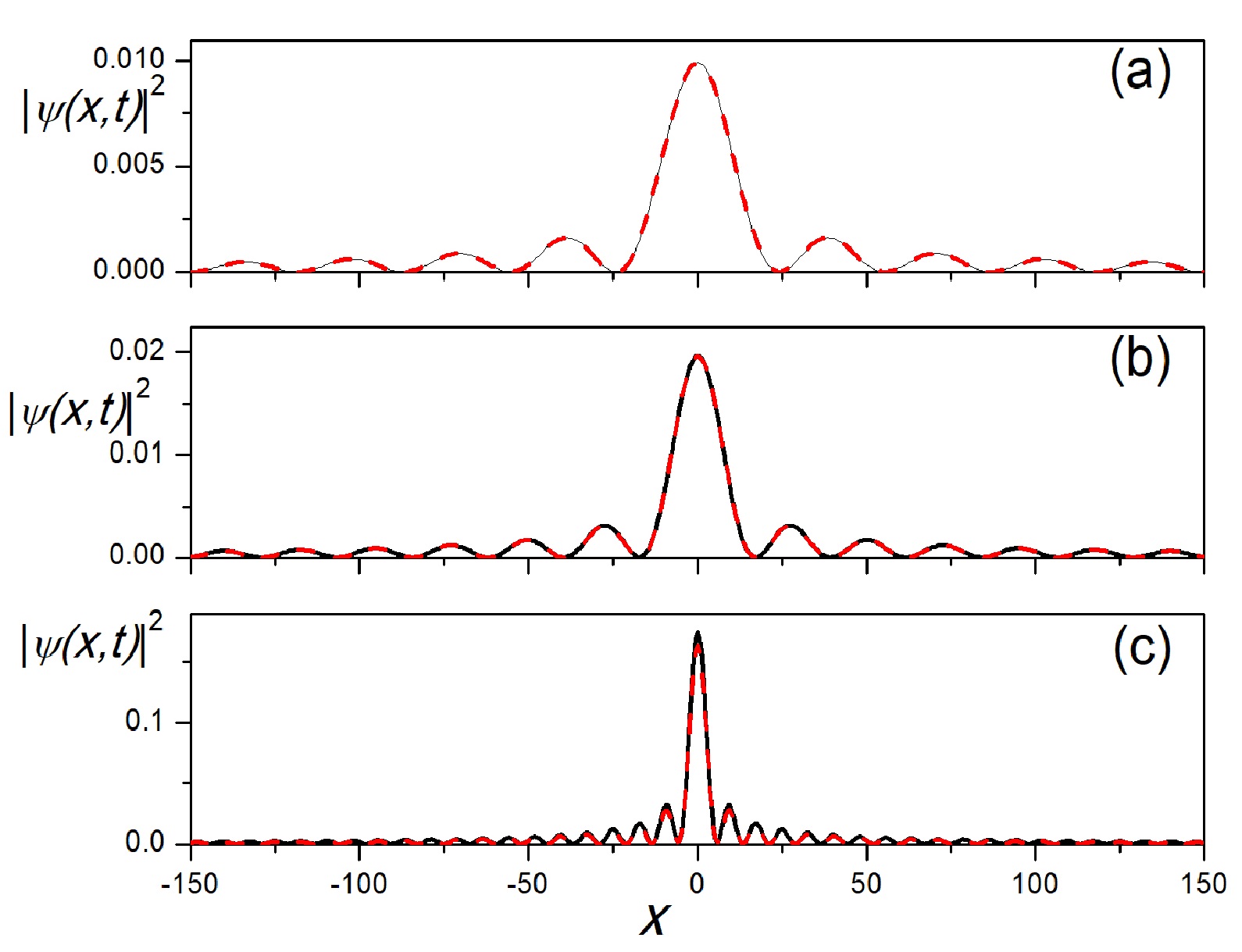} \caption{\label{fig4} Plot of the probability densities for a initial wave function given by a Bessel function $J_0(x)$ with parameters (a) $\alpha=10$, (b) $\alpha=5$ and (c) $\alpha=0.5$ for $t=5$ for the exact (solid lines) and first order approximate solutions (dashed lines). }
\end{figure}

We consider now the initial wave function given by a Bessel function \cite{Leija,Optica}
\begin{equation} 
\psi(x,0)=\exp \left( i\alpha x^2\right)  J_n(x),
\end{equation}
with $J_n(x)$ a Bessel function of  order $n$, defined as \cite{Arfken}
\begin{equation}
J_n(x)=\frac{1}{2\pi}\int_{-\pi}^{\pi} \exp\left( in\theta\right) \exp\left( -ix\sin\theta\right) d\theta.
\end{equation}
It is not difficult to show that the zeroth order solution is given by 
\begin{equation} 
\psi_0(x,t)=\frac{\exp\left[ if_1(t)x^2\right]} {\sqrt{1+2\alpha t}} J_n\left(\frac{x}{1+2\alpha t}\right),
\end{equation}
while the solution to first order reads
\begin{eqnarray}
& & \psi_1 \left( x,t\right)= \frac{\exp\left[ i f_1\left( t \right) x^2\right]}{\sqrt{1+2\alpha t}} \times
\nonumber \\
& & \left\lbrace 
\left[1+\frac{f_4\left( t \right) }{2}\right] 
J_n\left( \frac{x}{1+2\alpha t} \right)
-i \frac{f_4\left( t \right) }{4}
\left[ J_{n+2}\left( \frac{x}{1+2\alpha t} \right)+
J_{n-2}\left( \frac{x}{1+2\alpha t} \right)
\right]  
\right\rbrace. 
\end{eqnarray}
In order to show that the approximation is good, we write also the exact solution as
\begin{eqnarray} 
\psi(x,t)=\frac{ \exp\left[ if_1(t)x^2\right] }{2\pi\sqrt{1+2\alpha t}}\int_{-\pi}^{\pi} \exp\left( in\theta\right) \exp\left( -ix\sin\theta\right) \exp\left[ if_4(t)\sin^2\theta\right] d\theta,
\end{eqnarray}
which is a so-called generalized Bessel function \cite{Leija,Dattoli,Torre}. In Figure \ref{fig4}, we plot the probability densities for the exact (solid lines) and zeroth order solutions (dashed lines) which again show an excellent agreement.

\section{Conclusions}
We have shown that by adding a quadratic phase to an initial wave packet, its structure may be kept invariant through free evolution. The main result of this contribution is equation (\ref{invariant}), which shows clearly this fact. Although the invariance is  an approximation, it was  shown that it  perfectly matches the exact evolution. The price that has to be paid is the usual spread of the wave function due to free evolution, which is given here by the application of the squeeze operator to the initial wave function.\\

\end{document}